\documentclass[9pt]{extarticle}
\usepackage{amsmath,amssymb,graphicx, multirow}
\usepackage[margin=1in]{geometry}
\usepackage{xcolor}
\usepackage{fancyhdr}

\input{def.set}

\title{Collaborative Deep Learning for Speech Enhancement:\\A Run-Time Model Selection Method Using Autoencoders}

\author{Minje Kim\footnote{This work is copyrighted by the IEEE. Personal use of this material is permitted. However, permission to reprint/republish this material for advertising or promotional purposes or for creating new collective works for resale or redistribution to servers or lists, or to reuse any copyrighted component of this work in other works must be obtained from the IEEE.}\\
Indiana University Bloomington\\
School of Informatics and Computing\\
Department of Intelligent Systems Engineering\\
\texttt{minje@indiana.edu}}
\begin{document}

\maketitle

\thispagestyle{fancy} 
\begin{abstract}
We show that a Modular Neural Network (MNN) can combine various speech enhancement modules, each of which is a Deep Neural Network (DNN) specialized on a particular enhancement job. Differently from an ordinary ensemble technique that averages variations in models, the propose MNN selects the best module for the unseen test signal to produce a greedy ensemble. We see this as Collaborative Deep Learning (CDL), because it can reuse various already-trained DNN models without any further refining. In the proposed MNN selecting the best module during run time is challenging. To this end, we employ a speech AutoEncoder (AE) as an arbitrator, whose input and output are trained to be as similar as possible if its input is clean speech. Therefore, the AE can gauge the quality of the module-specific denoised result by seeing its AE reconstruction error, e.g. low error means that the module output is similar to clean speech. We propose an MNN structure with various modules that are specialized on dealing with a specific noise type, gender, and input Signal-to-Noise Ratio (SNR) value, and empirically prove that it almost always works better than an arbitrarily chosen DNN module and sometimes as good as an oracle result.
\end{abstract}
{\textbf{Keywords}: 
Speech Enhancement, Source Separation, Deep Learning, Modular Neural Networks, Autoencoders}

\section{Introduction}\label{sec:intro}

Deep learning has become one of the most popular frameworks for speech enhancement. The basic strategy of applying a Deep Neural Network (DNN) for the enhancement job is to learn a network that approximates the mapping function from a contaminated speech signal to its cleaned-up version. Various input and output features have been proposed. Xu et al. introduced a pre-training based Speech Denoising AutoEncoder (SDAE), which uses magnitudes of Fourier coefficients for both input and output \cite{XuY2014ieeespl}. Ideal Binary Masks (IBM) \cite{WilliamsonD2014jasa} and Ideal Ratio Masks (IRM) \cite{NarayananA2013icassp} are another common target representations. Huang et al.'s Deep Recurrent Neural Networks (DRNN) added the recurrent structure to SDAE, along with a discriminative term, too \cite{HuangP2015ieeeacmaslp}. More specialized speech features showed state-of-the-art performances such as cochleagrams \cite{WangY2013ieeeaslp}, Mel-Frequency Cepstrum Coefficients (MFCC)  \cite{NarayananA2013icassp}, and their combinations. Structural variations have been also investigated in the literature: deep clustering based on the independence of speakers \cite{HersheyJ2016icassp}, Long Short-Term Memory (LSTM) to handle long-term dependency of time-structured speech signals \cite{WeningerF2015lvaica}, deep unfolding networks to substitute the iterations in some estimation algorithms with a number of hidden layers \cite{LeRouxJ2015icassp}, etc. 

Although it is common to adapt the model for the unseen noise types in the dictionary-based approaches, in the deep learning-based models the adaptation largely relied on the generalization power of the already trained network. For example, in the semi-supervised source separation scenario, the system can learn the unseen noise dictionary from the noise source mixed in the test signal during run time along with the ordinary speech dictionary \cite{SmaragdisP2007ica, GermainF2015ieeespl, DuanZ2012lvaica}\footnote{Nonnegative Matrix Factorization (NMF) \cite{LeeDD99nature, LeeDD2000nips} and its variations are a common choice for the dictionary learning algorithm.}. As shown in \cite{GermainF2015ieeespl}, this semi-supervised technique is prone to overfitting due to the lack of the knowledge about the noise source. Meanwhile, Liu et al. performed some experiments to see the generalization power of a SDAE \cite{LiuD2014interspeech}. If the network was not exposed to a specific noise type during training, its performance degrades for the mixtures with that particular noise. Similar tests confirmed a suboptimal performance for unseen speakers and mixing weights as well. 

Recently, there have been DNNs that adapt to the unknown noise type by refining an already trained SDAE during run time. Kim and Smaragdis proposed an Adaptive SDAE (ASDAE) system, which is a vertical concatenation of two AEs: the bottom SDAE trained from known mixtures of speech and noise (to denoise them) and the top AE trained only from pure speech \cite{KimMJ2015lvaica}. It is based on the assumption that a properly trained AE for a source of interest (i.e. speech) can be used to judge the similarity between its input signal and the source, because the AE's reconstruction error will be low if the AE's input is speech as well, while for a non-speech signal the autoencoding performance is not guaranteed. Therefore, for a test mixture the bottom SDAE first estimates a cleaned-up version, which is subsequently fed to the top AE to calculate AE's reconstruction error as a measure of the denoising quality. Then, this reconstruction error is used to fine-tune the bottom SDAE through an additional backpropagation step. Another primitive refining scheme was proposed earlier by Williamson et al. \cite{WilliamsonD2014jasa}, where an NMF speech dictionary was used to further clean up the DNN results, although the use of NMF was limited to smoothing the results rather than fine-tune the main DNN. More recently, separable deep autoencoder showed promising denoising performance by having two AEs that model speech and noise separately. In there, the speech AE and an embedded NMF dictionary for an additional speech modeling are trained in advance, while the noise AE is trained from the test signal \cite{SunM2016ieeeacmaslp}. Those adaptive DNN models for speech denoising have focused only on adapting to unseen noise types. However, in practice we face a lot more variations such as in the ratio of sources' contributions, the frequency response of microphones, amount of reverberations, etc.

The proposed Modular Neural Network (MNN) assumes that it is easier to learn a smaller specialized DNN that work better for a particular enhancement job than a larger DNN for the general speech enhancement task as partially shown in \cite{KolbakM2017ieeeacmaslp}. Similarly, an MNN consists of local experts that provide various outputs for a test sample and a gating network that chooses the best module \cite{JacobsR1991nc}. The proposed MNN also includes the specialized speech enhancement modules as experts, while it uses the speech AE as its arbitrator. As the AE can be learned without any information about the participating modules, the proposed MNN is more scalable than the cases that need to learn the discriminative gating network for the selection. This can be seen as a Collaborative Deep Learning (CDL) system as well, because now we can invite any already-trained DNNs with different properties, and then the proposed selection scheme produces a greedy ensemble of them as the optimal result for the current test signal. The use of AE to determine the speech enhancement quality of another module is similar to the use in ASDAE, but the proposed MNN differs from ASDAE in that it accepts the best modular output instead of refining the modules so that it can prevent overfitting. 


\section{DNN for Supervised and Unsupervised Learning}

In this section we review two basic DNN systems for supervised speech denoising and unsupervised speech modeling, which are then combined to build the proposed system in Section \ref{sec:cdl}.

\subsection{DNN for Supervised Speech Denoising}

We start from a $D$-dimensional complex-valued Fourier spectrum at $t$-th time frame as an instantaneous mixture of a clean speech and noise spectra: $\bx_t=\bs_t+\bn_t$\footnote{From now on we drop the frame index for the notational convenience.}. Usually the input $\bx$ (sometimes along with its consecutive frames as well) goes through a feature extraction procedure to construct the input feature vector $\bar{\bx}\in\Real^{K^{(1)}}$. Now the goal of the training job is to learn the mapping function $\calF_{DNN}$ to produce an output vector $\by\in\Real^D$, which is either an estimation for the original speech features or a mask that can be later used to recover the speech. For the latter case, the feedforward and masking procedures work as follows:
\begin{equation}
\by=\calF_{DNN}(\bar{\bx}), \quad \hat{\bs}=\by\odot\bx,
\end{equation}
where $\odot$ stands for an element-wise multiplication and $\hat{\bs}$ is an estimation of ${\bs}$. For training, we can calculate the magnitude ratio as the masking vector $\bm=\frac{|\bs|}{|\bs+\bn|}$, and use them to prepare the training pairs $(\bar{\bx}, \bm)$. Hence, the training objective for a DNN with $L$ hidden layers is to minimize the sum of errors between the target masking vectors and the estimated ones:
\begin{equation}
    \argmin_{\bW^{(1)}, \cdots, \bW^{(L+1)}}\sum_t\calE\big(\bm_t\big\Vert\calF_{DNN}(\bar{\bx}_t)\big),
\end{equation}
where $\bW^{(l)}\in\Real^{K^{(l+1)}\times (K^{(l)}+1)}$ holds the network parameters at $l$-th layer, which participates in the feedforward procedure as follows:
\begin{align}
\nonumber\calF_{DNN}(\bar{\bx})&=\bz^{(L+2)}, \quad \bz^{(1)}=\bar{\bx},\\
\bz^{(l+1)}&=g^{(l)}\big(\bW^{(l)}\cdot[\big(\bz^{(l)}\big)^\top, 1]^\top\big).
\end{align}
Note that $\bz^{(l)}\in\Real^{K^{(l)}}$ is a vector of $K^{(l)}$ hidden unit outputs. There are a lot of choices for the activation function $g^{(l)}$, but the logistic function is commonly used for the last layer to ensure the soft masks between 0 and 1. Note also that the proposed model selection scheme works on any choice of the target representation of the participating DNNs if they can estimate a speech approximation $\hat{\bs}$. 

The mapping function $\calF_{DNN}$ could have been trained to perform well only on a subset of infinitely many mixing scenarios. For example, it can target on denoising only a particular person's noisy speech. On the other hand, the DNN might work for only a particular noise type, e.g. airplane noise. Finally, the DNN might have been trained only for a few choices of Signal-to-Noise Ratio (SNR) between the time domain signals $s$ and $n$ with sample index $\tau$, e.g. $SNR=10\log_{10} \frac{\sum_\tau s(\tau)^2}{\sum_\tau n(\tau)^2}$. In theory, there can be a very large and deep network that has been trained on all possible mixing cases. However, it is of our interest whether there is a systematic way to combine all the specialized models and to make the best out of them.

\subsection{Autoencoders for Unsupervised Speech Modeling}

AEs are another kind of neural networks whose target variables are set to be the same with the input, 
\begin{equation}\label{eq:ae_obj}
\calE\big(\bar{\bs}\big\Vert\calF_{AE}(\bar{\bs})\big).
\end{equation}
Therefore, a straightforward AE that models a source, e.g. speech, can be trained by using clean speech spectra for both input and target. Magnitudes of the complex-valued Fourier coefficients, $\bar{\bs}=|\bs|$, can serve as the features for our purpose.  

In the deep learning literature, a DAE has been also used to provide a greedy layer-wise feature learning \cite{VincentP2008icml, VincentP2010jmlr}, where the input vector goes through random perturbations such as masking noise: 
\begin{align}
&\tilde{\bs}=|\bs|\odot\mathbf{\nu}, \quad \mathbf{\nu}_i \sim \textrm{Bernoulli}(p),\\
\label{eq:dae_obj}&\calE\big(\bar{\bs}\big\Vert\calF_{DAE}(\tilde{\bs})\big),
\end{align}
with $p$ as the parameter for the Bernoulli distribution. Since the DAE has to produce the clean example from the corrupted inputs, the learned features are more robust and representative for the later use. Although the input and target are not exactly same, this kind of DAEs can still be seen as an unsupervised modeling because the corruption is done randomly without any supervision. 

A similar concept can be found in the dropout technique, too \cite{SrivastavaN2014jmlr}. During the feedforward process, dropout randomly turns off a certain number of units with the layer-wise Bernourlli random variable, $\nu^{(l)}\sim\textrm{Bernoulli}(p^{(l)})$, as its masking value:
\begin{equation}
\bz^{(l+1)}=g^{(l)}\big(\bW^{(l)}\cdot[\big(\mathbf{\nu}^{(l)}\odot\bz^{(l)}\big)^\top, 1]^\top\big),
\end{equation}
which is a procedure having a similar effect of averaging multiple thinned versions of the network. AEs with the dropout feature can also be seen as a DAE since not only their hidden units, but their input units are corrupted with masking noise. 

\textbf{A clarification for SDAE}: As we have reviewed in Section \ref{sec:intro}, SDAEs have been actively used to directly approximate the mapping from the contaminated speech to the clean ones in the context of supervised learning \cite{XuY2014ieeespl, LiuD2014interspeech}. For these supervised SDAEs, the objective is somewhat similar to that of an unsupervised AE in \eqref{eq:ae_obj} because their target variables are the clean speech features, too. However, it is different in the sense that it directly involves a few specific types of noise known in advance to perturb the input:
\begin{equation}\label{eq:sdae_obj}
\calE\big(\bar{\bs}\big\Vert\calF_{SDAE}(\bar{\bx}_t)\big).
\end{equation}
Hence, those SDAEs are not one of the unsupervised speech modeling techniques. Instead, it can serve as one of the participating enhancement modules in the proposed MNN system for CDL.  

\begin{figure}
    \centering
    \includegraphics[width=.5\columnwidth]{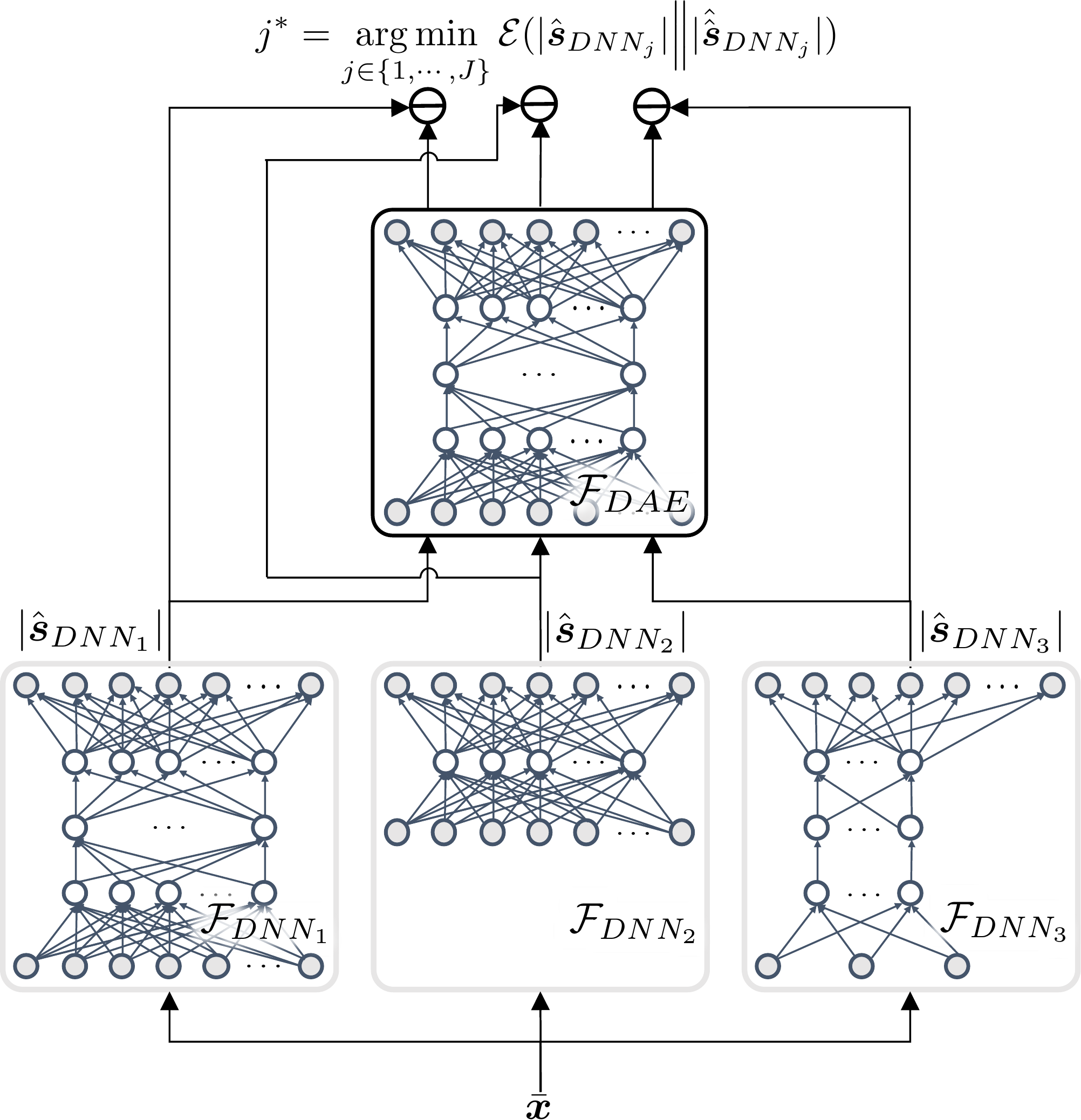}
    \caption{The proposed model selection procedure during run time}
    \label{fig:diagram}
\end{figure}

\section{The Proposed Modular Neural Network for Collaborative Deep Learning}\label{sec:cdl}

\subsection{The Proposed Architecture}

$\calF_{DNN_j}$ is one of the $J$ participating DNN modules in the MNN, which has been trained on only a subset of all the possible types of corruption. The structure of the modules can also vary in their number of layers and hidden units, choice of activation functions, use of recurrence and convolution, etc. Once each of them estimates a clean speech signal $\hat{s}_{DNN_j}$, it is fed to the model selector $\calF_{DAE}$ in the form of a magnitude spectrum, $|\hat{\bs}_{DNN_j}|$.

$\calF_{DAE}$ is trained in advance to produce a clean magnitude speech spectrum for its input of the same kind, while for the robustness to the various imperfection of $\hat{s}_{DNN_j}$ we choose to use a dropout-based DAE as in \eqref{eq:dae_obj} rather than an AE trained on clean speech. The key assumption is that a properly trained DAE will keep its clean speech spectra input intact, while its behavior for an unseen non-speech spectrum is not guaranteed. Consequently, the DAE error $\calE\big(\hat{\bs}_{DNN_j}\big\Vert\calF_{DAE}(|\hat{\bs}_{DNN_j}|\odot\nu^{(1)})\big)$ measures how much the input and speech are alike. Fig. \ref{fig:diagram} depicts this run-time process when $J=3$. $\hat{\hat{\bs}}_{DNN_j}=\calF_{DAE}(|\hat{\bs}_{DNN_j}|\odot\nu^{(1)})$ denotes the run-time DAE output. The final output of the proposed MNN system is the DNN module's output whose subsequent DAE error is the lowest: 
\begin{align}
\calF_{CDL}(\bar{\bx})&=\hat{s}_{DNN_{j^*}},\\
\label{eq:ae_err}j^*&=\argmin_{j\in\{1,\cdots,J\}} \calE(|\hat{\bs}_{DNN_j}|\Big\Vert|\hat{\hat{\bs}}_{DNN_j}|)
\end{align}
Alternatively, SNR can capture the discrepancy in time domain, too: 
\begin{equation}\label{eq:snr_err}
10\log_{10}\Big(\scriptsize{\sum_\tau} \hat{s}_{DNN_j}^2(\tau)\Big)\Big/\Big(\scriptsize{\sum_\tau} \big(\hat{s}_{DNN_j}(\tau)-\hat{\hat{s}}_{DNN_j}(\tau)\big)^2\Big).
\end{equation}

\subsection{Computational and Spatial Complexity}
The run-time computational and spatial complexity is clearly an issue with the proposed MNN for CDL method as every participating DNN needs to run a feedforward step. There are some promising approaches to compressing DNNs such as a low-rank approximation of the weight matrices \cite{XueJ2013interspeech} and networks that operate using bit logics and binary variables \cite{KimMJ2015icmlw, CourbariauxM2015nips}. Since the compressed networks claim their efficiency during run time, they can substitute the comprehensive ones for the model selection purpose. After the AE selection is done, we finally run the the best comprehensive DNN. We leave this network compression issue to future work.

\section{Experiments}

\subsection{The Speech AE}\label{sec:exp_ae}
Randomly chosen 400 utterances from the TIMIT training set are used for training (20 speakers $\times$ 2 genders $\times$ 10 utterances). Short-time Fourier transform with an 1024-point frame size and a 75\% overlap is used for the time-frequency conversion. Resilient backpropagation (Rprop) \cite{RiedmillerM1993icnn} technique is employed, and their parameters are found through a validation with additional four speakers: $0.5, 1.5, 10^{-7}$ and $10^{-1}$ for backtracking, acceleration, and minimum and maximum step sizes, respectively. We choose a modified Rectified Linear Unit (ReLU) as proposed in \cite{LiuD2014interspeech} for the activation function. Dropout parameters $p^{(l)}$ are all set to be $0.8$. The sum of the squared error is minimized during training. The batch size was set to be 1,000. They all converge in $5,000$ iterations.

Two DAEs with different network topologies model this speech data set: $\!\calF_{DAE_{128}}\!$ and $\!\calF_{DAE_{2048\times 2}}\!$. $\!\calF_{DAE_{128}}\!$ is with a single hidden layer of 128 units. A 513-dimensional magnitude spectrum works as its input and target. $\calF_{DAE_{2048\times 2}}$ is with two hidden layers, each of which has 2048 units. For $\!\calF_{DAE_{2048\times 2}}\!$, we concatenate three spectra $\big|[\bs_{t-1}^\top, \bs_{t}^\top, \bs_{t+1}^\top]^\top\big|$ to take the temporal dynamics of the signal into account, while its target is still a single spectrum, i.e. $|\bs_{t}|$. The two DAEs are compared to see if the larger and more complicated DAE measures the speech quality more correctly than the smaller one. 

\begin{table*}[t]
    \centering
    \resizebox{\textwidth}{!}{
    \begin{tabular}{|c|c|r|r|r|r||r|r||r|r||r|}
    \hline
        Final&\multirow{2}{*}{Test Noise} & \multicolumn{3}{c|}{Train Noise}  & \multirow{2}{*}{Chance} & \multicolumn{2}{c||}{SNR  \eqref{eq:snr_err}}   &   \multicolumn{2}{c||}{AE Recons. Error  \eqref{eq:ae_err}}   & \multirow{2}{*}{Oracle} \\
    \cline{3-5}\cline{7-10}
        Metric & &  Birds & Typing & Motorcycle & & $\calF_{DAE_{128}}$   & $\!\!\!\calF_{DAE_{2048\times 2}}\!\!\!$ & $\calF_{DAE_{128}}$  & $\!\!\!\calF_{DAE_{2048\times 2}}\!\!\!$  &  \\
    \hline
    \multirow{3}{*}{SDR} & Birds & 12.12 & 0.00 & 0.49 & 4.21 & 12.12 & 12.12 & 12.12 & 12.12 & 12.12\\
    & Typing & 0.18 & 12.74 & -0.83 & 4.03 & 12.74 & 12.74 & 12.74 & 12.74 & 12.74\\
    & Motorcycle & 5.80 & 3.40 & 10.15 & 6.45 & 9.93 & 9.81 & 9.93 & 9.81 & 10.15\\
    \hline
    \multirow{3}{*}{STOI} & Birds & 0.8501 & 0.7820 & 0.7636 & 0.7986 & 0.8501 & 0.8501 & 0.8501 & 0.8501 & 0.8501\\
    & Typing & 0.7130 & 0.8581 & 0.7161 & 0.7624 & 0.8581 & 0.8581 & 0.8581 & 0.8581 & 0.8581\\
    & Motorcycle & 0.7822 & 0.7737 & 0.8260 & 0.7940 & 0.8243 & 0.8229 & 0.8260 & 0.8229 & 0.8260 \\
    \hline
    \end{tabular}
    }
    \caption{Average SDR and STOI values of the final results chosen from three DNNs based on the proposed speech AE error.}
    \label{tab:noise}
\end{table*}

\begin{table*}[t]
    \centering
    \resizebox{\textwidth}{!}{
    \begin{tabular}{|c|c|r|r|r||r|r||r|r||r|}
    \hline
        Final&\multirow{2}{*}{Test Gender}  & \multicolumn{2}{c|}{Train Gender}  & \multirow{2}{*}{Chance} & \multicolumn{2}{c||}{SNR \eqref{eq:snr_err}}   &   \multicolumn{2}{c||}{AE Recons. Error \eqref{eq:ae_err}}   & \multirow{2}{*}{Oracle} \\
    \cline{3-4}\cline{6-9}
        Metric& &  Male & Female  & & $\calF_{DAE_{128}}$   & $\!\!\!\calF_{DAE_{2048\times 2}}\!\!\!$ & $\calF_{DAE_{128}}$  & $\!\!\!\calF_{DAE_{2048\times 2}}\!\!\!$  &  \\
    \hline
    \multirow{2}{*}{SDR} & Male & 10.38 & 8.15 & 9.27 & 9.78 & 10.10 & 9.65 & 10.06 & 10.45\\
    & Female & 7.58 & 11.00 & 9.29 & 10.85 & 10.91 & 10.76 & 10.79 & 11.02\\
    \hline
    \multirow{2}{*}{STOI} & Male & 0.8561 & 0.7951 & 0.8256 & 0.8362 & 0.8447 & 0.8338 & 0.8457 & 0.8561\\
    & Female & 0.8026 & 0.8503 & 0.8265 & 0.8477 & 0.8491 & 0.8467 & 0.8479 & 0.8505\\
    \hline
    \end{tabular}
    }
    \caption{The run-time selection results from two DNNs for two genders. }
    \label{tab:gender}
\end{table*}

\begin{table*}[t]
    \centering
    \resizebox{\textwidth}{!}{
    \begin{tabular}{|c|r|r|r|r|r||r|r||r|r||r|}
    \hline
        Final&\multirow{2}{*}{Test SNR} & \multicolumn{3}{c|}{Train SNR}  & \multirow{2}{*}{Chance} & \multicolumn{2}{c||}{SNR  \eqref{eq:snr_err}}   &   \multicolumn{2}{c||}{AE Recons. Error  \eqref{eq:ae_err}}   & \multirow{2}{*}{Oracle} \\
    \cline{3-5}\cline{7-10}
        Metric & &  -5 dB & 0 dB & +5 dB & & $\calF_{DAE_{128}}$   & $\!\!\!\calF_{DAE_{2048\times 2}}\!\!\!$ & $\calF_{DAE_{128}}$  & $\!\!\!\calF_{DAE_{2048\times 2}}\!\!\!$  &  \\
    \hline
    \multirow{3}{*}{SDR} & -5 dB & 6.89 & 7.00 & 6.37 & 6.75 & 7.01 & 6.94 & 7.03 & 7.01 & 7.27\\
    & 0 dB & 9.35 & 9.87 & 9.91 & 9.71 & 10.03 & 10.07 & 9.92 & 10.04 & 10.25\\
    & +5 dB & 11.55 & 12.24 & 12.79 & 12.19 & 12.64 & 12.80 & 12.49 & 12.65 & 12.90\\
    \hline
    \multirow{3}{*}{STOI} & -5 dB & 0.7569 & 0.7535 & 0.7380 & 0.7495 & 0.7494 & 0.7470 & 0.7508 & 0.7496 & 0.7609\\
    & 0 dB & 0.8253 & 0.8305 & 0.8268 & 0.8276 & 0.8289 & 0.8283 & 0.8289 & 0.8289 & 0.8340\\
    & +5 dB & 0.8775 & 0.8837 & 0.8863 & 0.8825 & 0.8856 & 0.8864 & 0.8850 & 0.8855 & 0.8883\\
    \hline
    \end{tabular}
    }
    \caption{Final results from DNNs that are dedicated to various input SNRs. }
    \label{tab:snr}
\end{table*}

\subsection{Experiment 1: Variations in the Noise Types}\label{sec:ex1}

For training we prepare 60 clean utterances per a noise type: (6 speakers) $\!\!\times\!\!$ (2 genders) $\!\!\times\!\!$ (5 utterances). They are then mixed up with one of three noise types chosen from \{``Birds", ``Typing", ``Motorcycle"\} \cite{DuanZ2012lvaica} at 0 dB SNR. We train one $512\!\times\!2$ DNN per one of the three noisy speech datasets as described in Section \ref{sec:exp_ae}, except some facts that (a) three input frames are always concatenated to form an input vector (b) the target is a masking vector of the center frame. The logistic function ensures the soft masking at the final layer. As for evaluation, we use both Signal-to-Distortion Ratio (SDR) \cite{VincentE2006ieeeaslp} and Short-Time Objective Intelligibility (STOI) \cite{TaalC2010icassp}. 

Three test datasets are from 20 gender-balanced clean utterances (5 from each of 4 speakers) mixed with different parts of the same three noise types. As shown in Table \ref{tab:noise}, if a DNN is trained and tested for the same kind of mixture, it performs the best: 12.12, 12.74 and 10.15 dB for the SDR and 0.8501, 0.8581, and 0.8260 for STOI. On the other hand, if we randomly choose one of the three trained systems at every time, the performance is a lot worse (the ``Chance" column). A truly optimal case is when we know the best module for each test sample (``Oracle" column), although in this experiment the system trained on the same noise type is always the best choice. 

Both selection metrics proposed in \eqref{eq:ae_err} and \eqref{eq:snr_err} assess the quality of all three participating DNNs' results, and then we average the SDR or STOI values of the selected results for the 20 test utterances. For all cases, the proposed method is better than chance. When one of the systems is absolutely better than the others (``Birds" and ``Typing") the proposed system reaches the oracle case. The shallow and deep AEs performs similarly in general, except the ``Motorcycle" case. We conjecture that a small AE is good enough when the participating DNNs are very specialized on a noise type like this.

\subsection{Experiment 2: Variations in Gender}

Next, we construct two datasets, each of which is from either 12 male or 12 female speakers. This time all ten noise types used in \cite{DuanZ2012lvaica} are mixed with the $12\times 5$ clean utterances, totalling $600$ per gender. Two gender-specific 2048$\times$2 DNNs are trained from these datasets, respectively. For testing we collect $10\times 5$ utterances per gender and mix them with the same ten noise types.

The module trained from the same gender performs better on the test set with the same gender than the wrong choice: 10.38 vs 8.15 and 11.00 vs 7.58 dB in SDR (Table \ref{tab:gender}), although their gap is smaller than Table \ref{tab:noise}. It might be because there can be a male test speaker whose voice is more similar to a female training speaker and vice versa. Similarly, the oracle case is better than the correct choice of DNN. Consequently, in this experiment the AEs' decision is not perfect, yet nearing the oracle case and showcasing much better results than chance. Note that $\calF_{{DAE}_{2048\times 2}}$ performs better than $\calF_{{DAE}_{128}}$.

\subsection{Experiment 3: Variations in the Input SNR}

For the final experiment, we randomly choose gender-balanced 12 speakers and their five utterances for training. Then, all ten noise types are mixed to build a set of 600 noisy utterances. For each set of 600 signals, we fix the loudness of the noise source to make the mixture has one of three SNR values, -5, 0, and +5. We train three 2048$\times$2 DNN modules on these. Table \ref{tab:snr} shows that the difference between the DNN modules is minute. For example, for the test signals with -5 dB SNR, the DNN system trained on 0 dB mixtures performs better than the correct choice in terms of SDR (7.00 vs 6.89 dB), because the correct DNN separated out the interfering noise too much, while introducing more artifacts which in turn decreased the overall separation quality. Yet, the DNN system trained on the -5 dB samples performs the best in terms of STOI. The proposed systems work better than chance most of the time (except the STOI value for -5 dB input case). $\calF_{{DAE}_{2048\times 2}}$ works better than $\calF_{{DAE}_{128}}$ in distinguishing the well denoised results that are only slightly different from each other (0 and +5 dB inputs). Note that this generic DNN with 0 dB mixture (9.87 dB) performs worse than the smaller noise-specific ones in Table \ref{tab:noise}, so it empirically shows that the correctly chosen specialized module outperforms the large generic network.

\section{Conclusion}

We proposed a collaborative deep learning method where multiple specialized DNN modules participate in the denoising job to produce various intermediate results. A DAE trained from clean speech judged the quality of the intermediate denoised results and chose the best one. The proposed MNN method showed better performance than the average of the candidate results in general. A shallow DAE was enough for most of the jobs, while the other deeper and larger DAE was more suitable for confusing high quality cases. The system was tested with variations in the noise type, gender, and input SNR. As future work, we plan to investigate more variations, e.g. reverberations and LSTMs for both DNN modules and DAEs.

\bibliographystyle{IEEEbib}

\bibliography{mjkim}

\begin{thebibliography}{10}

\bibitem{XuY2014ieeespl}
Y.~Xu, J.~Du, L.-R. Dai, and C.-H. Lee,
\newblock ``An experimental study on speech enhancement based on deep neural
  networks,''
\newblock {\em IEEE Signal Processing Letters}, vol. 21, no. 1, pp. 65--68,
  2014.

\bibitem{WilliamsonD2014jasa}
D.~S. Williamson, Y.~Wang, and D.~L. Wang,
\newblock ``Reconstruction techniques for improving the perceptual quality of
  binary masked speech,''
\newblock {\em Journal of the Acoustical Society of America}, vol. 136, pp.
  892--902, 2014.

\bibitem{NarayananA2013icassp}
A.~Narayanan and D.~L. Wang,
\newblock ``Ideal ratio mask estimation using deep neural networks for robust
  speech recognition,''
\newblock in {\em Proceedings of the IEEE International Conference on
  Acoustics, Speech, and Signal Processing (ICASSP)}, May 2013, pp. 7092--7096.

\bibitem{HuangP2015ieeeacmaslp}
P.~Huang, M.~Kim, M.~Hasegawa-Johnson, and P.~Smaragdis,
\newblock ``Joint optimization of masks and deep recurrent neural networks for
  monaural source separation,''
\newblock {\em IEEE/ACM Transactions on Audio, Speech, and Language
  Processing}, vol. 23, no. 12, pp. 2136--2147, Dec 2015.

\bibitem{WangY2013ieeeaslp}
Y.~Wang and D.~L. Wang,
\newblock ``Towards scaling up classification-based speech separation,''
\newblock {\em IEEE Transactions on Audio, Speech, and Language Processing},
  vol. 21, no. 7, pp. 1381--1390, July 2013.

\bibitem{HersheyJ2016icassp}
J.~R. Hershey, Z.~Chen, J.~{Le Roux}, and S.~Watanabe,
\newblock ``Deep clustering: Discriminative embeddings for segmentation and
  separation,''
\newblock in {\em Proceedings of the IEEE International Conference on
  Acoustics, Speech, and Signal Processing (ICASSP)}, Mar. 2016.

\bibitem{WeningerF2015lvaica}
F.~Weninger, H.~Erdogan, S.~Watanabe, E.~Vincent, J.~{Le Roux}, J.~R. Hershey,
  and B.~Schuller,
\newblock ``Speech enhancement with lstm recurrent neural networks and its
  application to noise-robust asr,''
\newblock in {\em Proceedings of the International Conference on Latent
  Variable Analysis and Signal Separation (LVA/ICA)}, Aug. 2015.

\bibitem{LeRouxJ2015icassp}
J.~{Le Roux}, J.~R. Hershey, and F.~Weninger,
\newblock ``Deep {NMF} for speech separation,''
\newblock in {\em Proceedings of the IEEE International Conference on
  Acoustics, Speech, and Signal Processing (ICASSP)}, Apr. 2015.

\bibitem{SmaragdisP2007ica}
P.~Smaragdis, B.~Raj, and M.~Shashanka,
\newblock ``Supervised and semi-supervised separation of sounds from
  single-channel mixtures,''
\newblock in {\em Proceedings of the International Conference on Independent
  Component Analysis and Blind Signal Separation (ICA)}, London, UK, 2007.

\bibitem{GermainF2015ieeespl}
F.~G. Germain and G.~J. Mysore,
\newblock ``Stopping criteria for non-negative matrix factorization based
  supervised and semi-supervised source separation,''
\newblock {\em IEEE Signal Processing Letters}, vol. 21, no. 10, pp.
  1284--1288, Oct 2014.

\bibitem{DuanZ2012lvaica}
Z.~Duan, G.~J. Mysore, and P.~Smaragdis,
\newblock ``Online {PLCA} for real-time semi-supervised source separation,''
\newblock in {\em Proceedings of the International Conference on Latent
  Variable Analysis and Signal Separation (LVA/ICA)}, 2012, pp. 34--41.

\bibitem{LeeDD99nature}
D.~D. Lee and H.~S. Seung,
\newblock ``Learning the parts of objects by non-negative matrix
  factorization,''
\newblock {\em Nature}, vol. 401, pp. 788--791, 1999.

\bibitem{LeeDD2000nips}
D.~D. Lee and H.~S. Seung,
\newblock ``Algorithms for non-negative matrix factorization,''
\newblock in {\em Advances in Neural Information Processing Systems (NIPS)}.
  2001, vol.~13, MIT Press.

\bibitem{LiuD2014interspeech}
D.~Liu, P.~Smaragdis, and M.~Kim,
\newblock ``Experiments on deep learning for speech denoising,''
\newblock in {\em Proceedings of the Annual Conference of the International
  Speech Communication Association (Interspeech)}, Sep 2014.

\bibitem{KimMJ2015lvaica}
M.~Kim and P.~Smaragdis,
\newblock ``Adaptive denoising autoencoders: A fine-tuning scheme to learn from
  test mixtures,''
\newblock in {\em Proceedings of the International Conference on Latent
  Variable Analysis and Signal Separation (LVA/ICA)}, August 2015.

\bibitem{SunM2016ieeeacmaslp}
M.~Sun, X.~Zhang, H.~Van hamme, and T.~F. Zheng,
\newblock ``Unseen noise estimation using separable deep auto encoder for
  speech enhancement,''
\newblock {\em IEEE/ACM Transactions on Audio, Speech, and Language
  Processing}, vol. 24, no. 1, pp. 93--104, Jan 2016.

\bibitem{KolbakM2017ieeeacmaslp}
M.~Kolb{\ae}k, Z.~H. Tan, and J.~Jensen,
\newblock ``Speech intelligibility potential of general and specialized deep
  neural network based speech enhancement systems,''
\newblock {\em IEEE/ACM Transactions on Audio, Speech, and Language
  Processing}, vol. 25, no. 1, pp. 153--167, Jan 2017.

\bibitem{JacobsR1991nc}
R.~A. Jacobs, M.~I. Jordan, S.~J. Nowlan, and G.~E. Hinton,
\newblock ``Adaptive mixtures of local experts,''
\newblock {\em Neural Computation}, vol. 3, no. 1, pp. 79--87, Mar. 1991.

\bibitem{VincentP2008icml}
P.~Vincent, H.~Larochelle, Y.~Bengio, and P.-A. Manzagol,
\newblock ``Extracting and composing robust features with denoising
  autoencoders,''
\newblock in {\em Proceedings of the International Conference on Machine
  Learning (ICML)}, 2008, pp. 1096--1103.

\bibitem{VincentP2010jmlr}
P.~Vincent, H.~Larochelle, I.~Lajoie, Y.~Bengio, and P.-A. Manzagol,
\newblock ``Stacked denoising autoencoders: Learning useful representations in
  a deep network with a local denoising criterion,''
\newblock {\em Journal of Machine Learning Research}, vol. 11, pp. 3371--3408,
  Dec. 2010.

\bibitem{SrivastavaN2014jmlr}
N.~Srivastava, G.~Hinton, A.~Krizhevsky, I.~Sutskever, and R.~Salakhutdinov,
\newblock ``Dropout: A simple way to prevent neural networks from
  overfitting,''
\newblock {\em Journal of Machine Learning Research}, vol. 15, no. 1, pp.
  1929--1958, Jan. 2014.

\bibitem{XueJ2013interspeech}
J.~Xue, J.~Li, and Y.~Gong,
\newblock ``Restructuring of deep neural network acoustic models with singular
  value decomposition.,''
\newblock in {\em Proceedings of the Annual Conference of the International
  Speech Communication Association (Interspeech)}, 2013, pp. 2365--2369.

\bibitem{KimMJ2015icmlw}
M.~Kim and P.~Smaragdis,
\newblock ``Bitwise neural networks,''
\newblock in {\em International Conference on Machine Learning (ICML) Workshop
  on Resource-Efficient Machine Learning}, Jul 2015.

\bibitem{CourbariauxM2015nips}
M.~Courbariaux, Y.~Bengio, and J.-P. David,
\newblock ``Binaryconnect: Training deep neural networks with binary weights
  during propagations,''
\newblock in {\em Advances in Neural Information Processing Systems (NIPS)},
  2015, pp. 3105--3113.

\bibitem{RiedmillerM1993icnn}
M.~Riedmiller and H.~Braun,
\newblock ``A direct adaptive method for faster backpropagation learning: the
  rprop algorithm,''
\newblock in {\em IEEE International Conference on Neural Networks}, 1993, pp.
  586--591 vol.1.

\bibitem{VincentE2006ieeeaslp}
E.~Vincent, C.~Fevotte, and R.~Gribonval,
\newblock ``Performance measurement in blind audio source separation,''
\newblock {\em IEEE Transactions on Audio, Speech, and Language Processing},
  vol. 14, no. 4, pp. 1462--1469, 2006.

\bibitem{TaalC2010icassp}
C.~H. Taal, R.~C. Hendriks, R.~Heusdens, and J.~Jensen,
\newblock ``A short-time objective intelligibility measure for time-frequency
  weighted noisy speech,''
\newblock in {\em Proceedings of the IEEE International Conference on
  Acoustics, Speech, and Signal Processing (ICASSP)}, 2010.

\end{thebibliography}

\end{document}